# Generation of Patient-specific Structured Hexahedral Mesh of Aortic Aneurysm Wall

**F. Alkhatib, G. C. Bourantas, A. Wittek, and K. Miller**

**Abstract** Abdominal Aortic Aneurysm (AAA) is an enlargement in the lower part of the main artery "Aorta" by 1.5 times its normal diameter. AAA can cause death if rupture occurs. Elective surgeries are recommended to prevent rupture based on geometrical measurements of AAA diameter and diameter growth rate. Reliability of these geometric parameters to predict the AAA rupture risk has been questioned, and biomechanical assessment has been proposed to distinguish between patients with high and low risk of rupture. Stress in aneurysm wall is the main variable of interest in such assessment. Most studies use finite element method to compute AAA stress. This requires discretising patient-specific geometry (aneurysm wall and intraluminal thrombus ILT) into finite elements/meshes. Tetrahedral elements are most commonly used as they can be generated in seemingly automated and effortless way. In practice, however, due to complex aneurysm geometry, the process tends to require time consuming mesh optimisation to ensure sufficiently high quality of tetrahedral elements. Furthermore, ensuring solution convergence requires large number of tetrahedral elements, which leads to long computation times. In this study, we focus on generation of hexahedral meshes as they are known to provide converged solution for smaller number of elements than tetrahedral meshes. We limit our investigation to already existing algorithms and software packages for mesh generation. Generation of hexahedral meshes for continua with complex/irregular geometry, such as aneurysms, requires analyst interaction. We propose a procedure for generating high quality patient-specific hexahedral discretisation of aneurysm wall using the algorithms available in commercial software package for mesh generation. For aneurysm cases, we demonstrate that the procedure facilitates patient-specific mesh generation within timeframe consistent with clinical workflow constraints while requiring only limited input from the analyst. As a future aspect, the methodology used in this article can be automated with minimal user interaction, and further to develop a robust algorithm for structured mesh generation in aneurysm wall.

F. Alkhatib (✉) · G. C. Bourantas · A. Wittek · K. Miller
Intelligent Systems for Medicine Laboratory, The University of Western Australia, Perth, Western Australia, Australia
E-mail: farah.alkhatib@research.uwa.edu.au

K. Miller
Harvard Medical School, Boston, MA, USA

## 1    Introduction

Abdominal Aortic Aneurysm (AAA) is a chronic vascular disease of elderly men (over 65 years old), while women has less chance by 4-6 times in same age group [1]. It is a permanent and irreversible enlargement in the lower part of the aorta [2], the main artery that pumps blood from the heart to the rest of human body. AAA is usually diagnosed incidentally by unrelated follow-up or examination as it is symptomless disease [3]. The most fatal event of AAA is rupture, where mortality rate can reach up to 90% [4] leading to 200,000 deaths annually worldwide [5] and around 1,500 deaths yearly in Australia and the Oceania region [6].

According to current AAA management, patients undergo elective surgical intervention if their maximum aortic diameter is more than 55 mm for men and 50 mm for women [7, 8], while below these recommended thresholds patients are placed on surveillance program that monitors the aneurysm growth rate. The surgery is recommended if growth rate exceeds 10 mm/year. Australia has a high rate of AAA repairs below these recommended thresholds compared to other Western countries. However, the probability for aneurysms with aortic diameter of $40-50$ mm under surveillance to rupture is only 0.4% per year, which is lower than the risk of death due to the postoperative complications [3].

This raises the question of how to best manage AAAs as there is a balance between interventions to prevent AAA rupture versus overtreatment that may cause harm to patients and incur non-essential medical cost. Over the last 25 years, researchers introduced different AAA biomechanical rupture risk indicators or indices to identify patients at high risk of AAA rupture [9-13] and conversely those at low risk for whom surgical intervention can be avoided. Aneurysm wall stress calculation is the main non-invasive biomechanical assessment used for these rupture indices, in addition to population-based strength statistical models used for some of rupture indices [9]. Evaluation of such indices is beyond the scope of this study. Finite element method is widely used for AAA stress calculations that requires discretising patient-specific geometry (aneurysm wall and intraluminal thrombus ILT) into finite elements/meshes as a part of creating the finite element model.

Tetrahedral mesh generation for patient-specific AAA geometries is used in computational biomechanical analysis as it is believed it can be created automatically with high element quality without expertise in computational grid generation [13, 14]. In a study done by Miller et al. [15], AAA (aneurysm wall and ILT) finite element models contained more than 1 million tetrahedral elements. This high number of elements ensures convergent solution, but tends to result in relatively long computational times. Furthermore, automated elimination of low quality tetrahedral elements typically requires application of mesh optimisation procedures. From our experience, presence of even small number of low quality elements may lead to unreasonably long optimisation times (up to around $40-50$ minutes of a personal computer with Intel quad-core i7 processs). Therefore, we focus on hexahedral meshes as they require smaller number of elements than tetrahedral meshes [16]. For aneurysm walls discretised using 30,000 to 50,000 hexahedral elements, around 500,000 tetrahedral elements were needed to achieve similar geometric discreitsation accuracy [17].

Generation of high quality structured (mapped) hexahedral finite element meshes of healthy blood vessel walls can be done automatically by defining the vessels centerlines [18-20] using freely available software such as pyFormex (https://github.com/dladd/pyFormex) and Gmsh (https://gmsh.info/). This, however, does not extent to complex/irregular geometry of AAAs. Generation of structured hexahedral meshes of AAAs tends to require expert's knowledge of finite element meshing procedures and substantial manual effort of the analyst. Specialised mesh generation code developed by Tarjuelo-Gutierrez et al. [21] facilitates construction of hexahedral meshes for aneurysm wall and thrombus including the bifurcations. It relies on connecting the extracted axial and longitudinal lines in the aneurysm from the manual MRI (magnetic resonance imaging) segmentation and the calculated aneurysm centerline. Need for substantial effort of the analyst was also reported in the studies using well established commercial mesh generators. Application of CEM CFD 14.5 (Ansys Inc., USA) to create patient-pecific hexahedral finite element meshes of aneurysm wall required 4 – 8 hours of analyst's work per case [12, 22]. Mayr et al. [23] used CUBIT mesh generator (https://cubit.sandia.gov/) to create hexahedral elements for aortic aneurysms to be used in fluid-structure interaction simulation. Distinct advantage of CUBIT that it can automatically partition complex geometries into mappable volumes to build structured hexahedral mesh. CUBIT is available for US government use only. However, its commercial version, Coreform Cubit (https://coreform.com/products/coreform-cubit/free-meshing-software/), has no such restriction. A4Clinics Vascops (http://www.vascops.com/en/vascops-A4clinics.html) software to biomechanically analyse AAA rupture risk creates a hexahedral aneurysm wall with a minimal user interaction. Their meshing algorithm limits any mesh refinement along the circumferential and axial directions of aneurysm wall and ILT, which creates one layer through wall thickness and coarse elements for thick ILT [24]. However, in several studies it has been argued that at least two elements across the AAA wall thickness is needed for convergred solution in terms of stress computation [25] . Y. Zhang et al. [16, 26] have successfully created unstructured hexahedral meshes from volumetric data (medical images as an example) to be used in FEA. Automated unstructured hexahedral elements for aneurysm wall and ILT using Harpoon (http://www.sharc.co.uk/index.htm) was done by Maier et al. [27]. Our experience indicates that unstructured hexahedral meshes may contain some poor quality elements, in particular elements with very poor (close to zero or even negative) Jacobian quality measure.

In this article, we demonstrate a procedure to create a high quality patient-specific structured hexahedral mesh of aortic aneurysm wall models using commercially available mesh generators for stress computation in the aneurysam wall. We use tetrahedral elements for the intraluminal thrombus (ILT) because of its complex geometry. In addition, accurate ILT stress analysis is not a variable of interest as an indicator of AAA rupture risk, and hence we use tetrahedral elements.

## 2    Methods

### 2.1  Patient's data and patient-specific AAA geometry

A contrast-enhanced computed tomography angiography (CTA) image data-set of four abdominal aortic aneurysm (AAA) patients with an average maximum aortic diameter of 55 mm (standard deviation = 9 mm) were used to demonstrate the meshing techniques proposed and used in this work. The CTA images were acquired at Fiona Stanley Hospital (Murdoch, Western Australia, Australia) using SOMATOM Definition Flash CT Scanner (Siemens Healthineers AG, Forchheim, Germany). The spatial resolution (voxel size) of the CTA images is 0.625x0.625x1.5 mm$^3$. Patient's gave their informed consent before acquiring the images according to the Declaration of Helsinki.

The patient-specific AAA geometries were segmented from the CTA images using the open-source medical image analysis package, 3D Slicer (https://www.slicer.org/) [28]. The contrast-enhanced images allowed an automated segmentation for the lumen (blood channel) using the threshold algorithm in the segmentation module. The Aneurysm (wall and the intraluminal thrombus 'ILT') needed some manual work to distinguish between the aneurysm and surrounding tissues. Figure 1 shows the segmented patient-specific AAA geometry for a selected case (Patient 1), the blue geometry is the aneurysm wall and the red geometry is the ILT. We assumed constant wall thickness of 1.5 mm for the aneurysm wall, as there is no reliable method to accurately determine AAA wall thickness from CTAs only has developed yet [29].

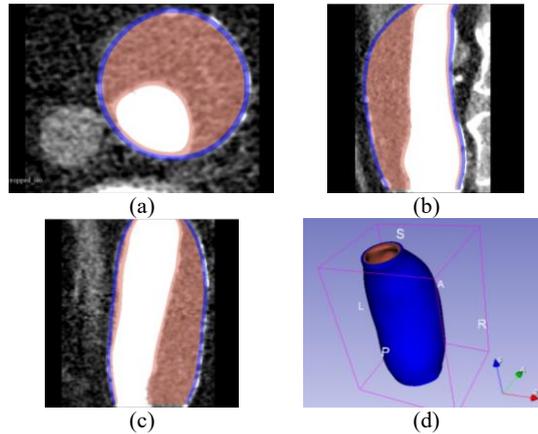

(a)

(b)

(c)

(d)

**Fig. 1** Patient-specific abdominal aortic aneurysm (AAA) geometry segmented from computed tomography angiography (CTA) using 3D Slicer. The segmented aneurysm wall with constant thickness of 1.5 mm is shown in blue and the segmented intraluminal thrombus (ILT) is shown in red; (a) a slice from the axial view of AAA, (b) a slice from the sagittal view of AAA, (c) a slice from the coronal view of AAA, and (d) the 3D rendered AAA.

## *2.2 Generation of patient-specific AAA computational grids*

### 2.2.1 Patient-specific AAA meshes

Different meshing tools and algorithms can be used to generate hexahedral elements for the aneurysm wall, we mentioned some of them in the introduction. We tried a simple in-house MATLAB code that creates the aneurysm topology using the aortic centreline and maximum distances from this centreline to the aortic wall. This algorithm defines subdivided circles and ellipses orthogonal to the centreline ready to be connected using splines to form surface quadrilateral meshes. We faced two main issues in this early-stage code, (1) these circles and ellipses may overlap in the locations that have big change in wall curvature, and (2) a smoothing technique (Laplace smoothing as an example [30]) should be used to improve the elements shape and mesh quality.

Fully automated hexahedral meshing was not possible for the aneurysm wall using available open-source and commercial mesh generators because of its irregular and asymmetrical shape. We used the mesh generator available in ABAQUS/CAE (https://www.3ds.com/products-services/simulia/products/abaqus/) finite elemen pre-processor. It provides high quality element generation although it strongly relies on user's expertise and requires substantial input (manual mesh generation work) from the user.. The geometry needs to be subdivided into many partitions in order to create the structured hexahedral mesh. We used the commercial mesh generation software Altair HyperMesh (https://www.altair.com/hypermesh) to create a high quality hexahedral aneurysm wall models, as the ISML (Intelligent Systems for Medicine Laboratory) team has many years of experience in using it, and it is a powerful finite element meshing tool that generates high quality elements from CAD (Computer-aided design) or image-based geometries. It should be noted that any mesh generator can be used. In the following text we describe the procedure used in HyperMesh to create the meshes.

We imported patient-specific geometry of AAA wall extracted from the CTA images previously in STL format (Stereo Lithography file format). Because of the irregular geometry of AAA, the geometry should be partitioned to create the structured (mapped meshing) hexahedral mesh. Figure 2 shows the four aneurysm wall geometries meshed in this study.

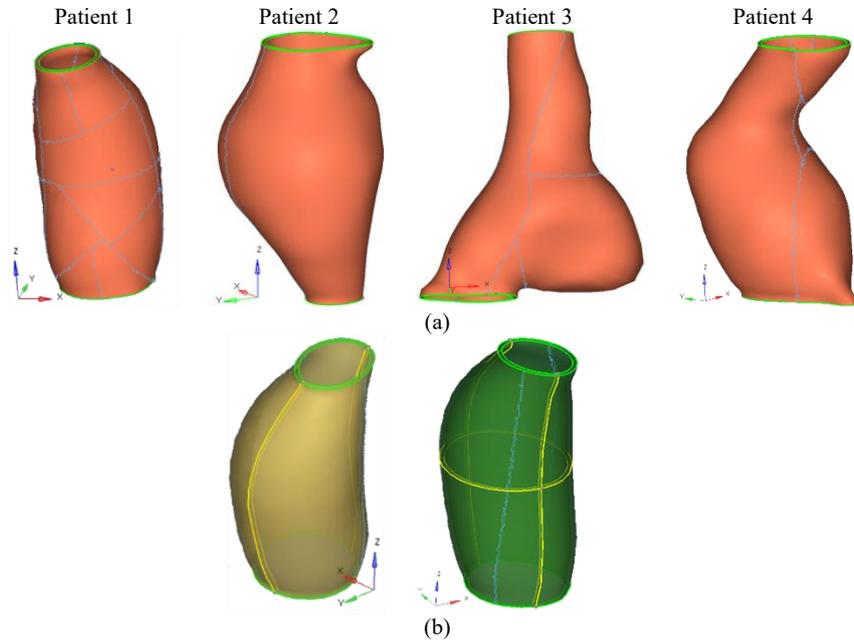

**Fig. 2** Aneurysm wall geometries extracted from CTA images and imported in STL format to HyperMesh; (a) geometries of the four patients, orange geometries cannot be meshed directly and at least one partition is needed to create the mapped mesh, (b) partitioned geometry ready for mapped (structured) meshing (Patient 1) using one plane (yellow geometry) or two planes (green geometry), the yellow lines present the planes used as partitions. According to HyperMesh colour code, yellow geometries can be meshed in one direction and green geometries in three directions (from three sides).

The 3D hexahedral mesh (Figure 3a) was created by sweeping the 2D meshed ring (Figure 3b) along the aneurysm wall. The created 2D top ring of quadrilateral elements in the aneurysm wall defines the element size and number of hexahedral layers through wall thickness. We used two elements through wall thickness with an element size of 0.75 mm.

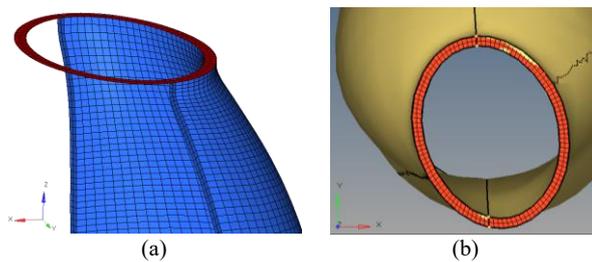

**Fig. 3** Hexahedral meshing of aneurysm wall, (a) section of aneurysm wall (blue elements) showing the hexahedral meshes created by sweeping the quadrilateral 2D ring (red elements), and (b) top view of the aneurysm wall geometry that has the generated 2D top ring of quadrilateral elements used to create the 3D volume wall (hexahedral elements), element size is 0.75 mm.

We created tetrahedral elements for the aneurysm intraluminal thrombus (ILT) using HyperMesh because of its complex shape. The transition between the quadrilateral surface meshes to tetrahedral volume meshes was important to create the shared conformal surface between the hexahedral aneurysm wall and the tetrahedral ILT (Figure 4a). We used the inner's wall surface topology (nodes location) generated previously to create the outer surface of the ILT (triangular elements). We imported the inner surface of the ILT (the surface of the lumen 'blood channel' segmented in 3D Slicer earlier) and created a dummy mesh to close the top and bottom caps of the ILT (Figure 4b).

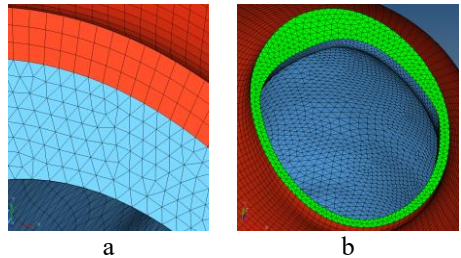

a       b

**Fig. 4** Generation of tetrahedral intraluminal thrombus (ILT) using HyperMesh, (a) part of the top view of abdominal aortic aneurysm (AAA) showing the wall in red and ILT in blue, (b) 2D top cap created (green) to close the volume of ILT and generate the tetrahedral volume mesh.

As the ILT volume is closed, a tetrahedral automated mesh filled this empty volume, allowing the algorithm to reduce number of elements inside ILT, and hence the computation time in the finite element solution is minimised. We kept the existing nodes/locations for both surfaces of wall and ILT, and allowed the splitting of quadrilaterals into triangles to avoid the creation pyramid elements. The upper and lower caps created (dummy mesh) we set them as freely adjusted nodes. A section view of the tetrahedral ILT created is shown in Figure 5.

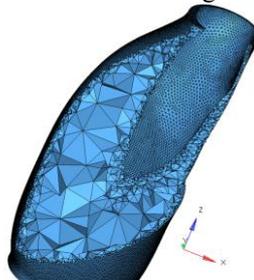

**Fig. 5** Meshed tetrahedral intraluminal thrombus (ILT).

### 2.2.2 Element quality

For hexahedral elements, we used two measures to check elements quality: (1) normalised/scaled Jacobian quality measure [26, 31], a Jacobian value of 0.6 and higher is recommended [32], and (2) the minimum and maximum allowable interior angles of a quadrilateral face, the suggested limits are between 45 and 135 degrees [32]. Both measures are important as we found that some elements have high Jacobian with maximum or minimum interior angles that are out of the recommended range. Figure 6 shows and describes an example of low quality element. From our experience, we confirmed that more partitions or sections in the geometry can solve such problem. Additionally, decreasing the element size may also eliminate the creation of low quality elements.

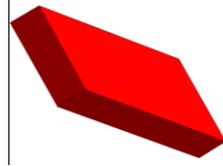

**Fig. 6** Example of low quality hexahedral element with Jacobian of 0.95, but with minimum interior angle of 39.25° and maximum interior angle of 141.36° that are out of the recommended allowable angles (45° - 135°) [32].

We checked the tetrahedral elements quality using two measures: (1) maximum and minimum allowable interior angle for triangles, and (2) the volumetric skew. The recommended interior angles ranges between 30 and 120 degrees [32]. The volumetric skew is 1 minus the ratio of the actual tetrahedron volume to an equilateral tetrahedron volume of same circumradius (circumradius is the radius of a sphere passing through the four vertices of the tetrahedron). A value of 1 means a flat tetrahedral element as seen in Figure 7. We found that all poor quality elements are located at the top or bottom edges of the ILT which are areas of no interest.

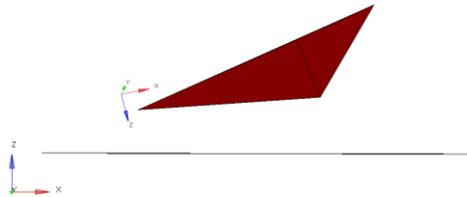

**Fig. 7** Two views showing a bad quality tetrahedral element with volumetric skew of 1 which means a flat tetrahedral element.

### 2.3 Stress computation in AAA wall

We used linear static finite element analysis implemented in ABAQUS/Standard finite element code [33] (https://www.3ds.com/products-services/simulia/products/abaqus/) to calculate the aneurysm wall stress, based on the previous study by Joldes et al. [34] that compared direct non-linear and inverse non-linear finite element procedures [35, 36] with their elastic linear approach, they concluded that the deformed AAA geometry and patient's blood pressure are the only inputs needed for acceptable AAA stress calculations. Furthermore, their study showed that stress calculation in AAA is independent of material properties of aortic tissue, as stress is calculated by balancing the internal forces of the aorta with the applied pressure (patient's blood pressure) [13, 34, 37].

We defined an incompressible linear material model for the aneurysm wall, while the aneurysm ILT incompressible linear material model was more compliant by 20 times. We used the hybrid 20-noded quadratic element C3D20RH for the hexahedral elements, this reduced integration element has 8 integration element, and the hybrid 10-noded quadratic element C3D10H for the tetrahedral elements. The hybrid formulation prevents volumetric locking for these nearly incompressible materials.

The aneurysm was uniformly loaded at the internal surface of the ILT by the patient-specific blood pressure measured 5 minutes before acquiring the scan. We used mean arterial blood pressure (MAP) that is calculated from the systolic and diastolic pressures (MAP = 1/3 systolic pressure + 2/3 diastolic pressure). The aneurysm was fully supported at its top and bottom edges.

We analysed the maximum principal stress of the patient-specific aneurysm walls as it represent the internal forces that balance the blood pressure in the blood vessel [13]. The residual stresses of the aorta were not taken into account in this study, as we focus on the method of generating high quality hexahedral meshes for aneurysm wall rather than the AAA rupture assessment through stress computation. We compared the stress results of hexahedral walls with finite element models of same patients that contained tetrahedral meshes for the aneurysm walls. The tetrahedral models were created using Gmsh used within from BioPARR (https://bioparr.mech.uwa.edu.au/) and stress calculations using ABAQUS/Standard. BioPARR is freely-available open-source software for utilising finite element method for AAA stress computation [13]. In addition to maximum principal stress, we analyse the 99th percentile of maximum principal stress in aneurysm wall [38] to eliminate any artefacts and uncertainties created during AAA geometry segmentation or generation of the computational grids.

# 3 Results

## 3.1 Convergence suggests non-linear analysis

We performed a mesh convergence study on one of the analysed AAAs (Patient 1) to ensure that computation of stress in aneurysm wall is independent of mesh size while reducing computational time. We created three hexahedral aneurysm wall models. The first model included 2 layers of hexahedral elements through wall thickness, the second model had 3 layers while the third model had 4 layers. We calculated the maximum principal stress in those three models using ABAQUS/Standard on Intel(R) Core(TM) i7-5930K CPU @ 3.50 GHz with 64.0 GB of RAM running Windows 8 OS. The finite element models did not include

ILT. The inner surface of the hexahedral wall was loaded with 12 kPa (MAP, mean arterial pressure).

Table 1 summarises the mesh characteristics of each studied finite element model. It also includes the peak and 99th percentile values of maximum principal stress for the three models. The peak stress values usually occur at the fixed nodes of the top and bottom edges of the aneurysm wall, thus we compared 99th percentile of stress.

**Table 1** Mesh characteristics, maximum principal stress values (peak and 99th percentile), and finite element models computation time for the studied models.

|  | Model 1 | Model 2 | Model 3 |
|---|---|---|---|
| No. of hexahedral elements through wall thickness | 2 | 3 | 4 |
| Element size (mm) | 0.75 | 0.5 | 0.375 |
| No. of elements | 27,972 | 95,190 | 225,676 |
| No. of nodes | 154,734 | 477,787 | 1,075,083 |
| Peak of maximum principal stress (MPa) | 0.5300 | 0.6070 | 0.6831 |
| 99th percentile of maximum principal stress (MPa) | 0.2700 | 0.2620 | 0.2557 |
| Computation time (sec) | 23 | 103 | 568 |

We also selected four nodes in the three aneurysm wall models to compare their maximum principal stress results. The selected nodes were in the middle of the aneurysm wall, two of them were on the outer surface of the wall and two on the inner surface. The four nodes had the exact location/coordinates for the three studied aneurysm walls. Figure 8 shows the convergence of the 4 selected nodes.

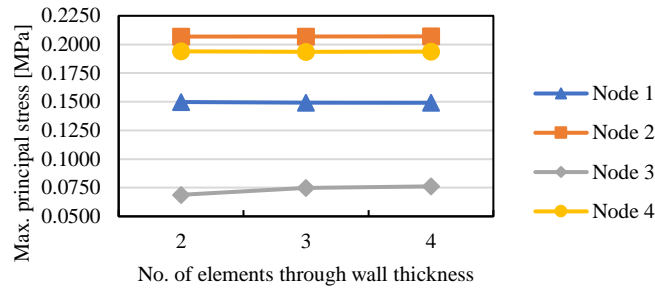

**Fig. 8** Convergence of the selected 4 nodes in the three studied meshed models with respect to their maximum principal stress values.

Figure 9 shows a quantitative representation of maximum principal stress for the three studied models with respect of a percentile rank that we call maximum principal stress percentile plot. This plot shows a uniform distribution of maximum principal stress values over all the nodes in the finite element models. The three models shows a perfect match regarding the stress distribution, thus we conclude that 2 layers of hexahedral elements through wall thickness are sufficient for a converged solution at the same time the computation time is hugely decreased compared to the other 2 models.

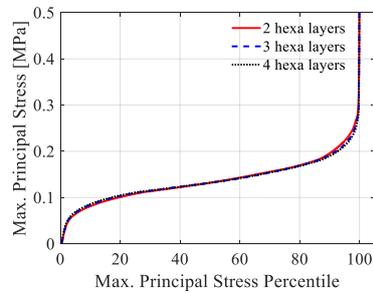

**Fig. 9** Maximum principal stress percentile plot for the three studied models to show the mesh independency for stress calculations.

### 3.2 Computational grids and element quality

The created computational grids (meshes) are depicted in Figure 10, we used 2 hexahedral elements through wall thickness and an element size of 0.75 mm.

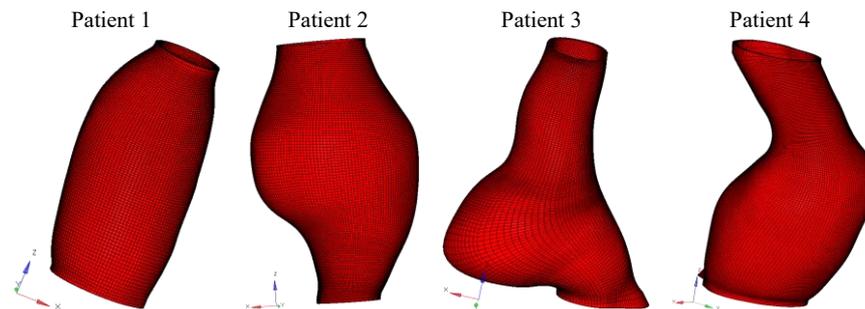

**Fig. 10** The hexahedral aneurysm walls created using HyperMesh with an element size of 0.75 mm.

Table 2 summarises number of low quality elements according to the quality measures used, Jacobian and minimum/maximum interior allowable angles for hexahedral elements, and volumetric skew and minimum/maximum interior allowable angles for tetrahedral elements. We noticed that low quality tetrahedral elements in ILT are located at the top and bottom edges of the aneurysm, where it is fully supported, and form very small number compared to the total number of ILT elements in AAA which does not exceed 0.02% of the total elements number in the AAA model.

**Table 2** Summary of poor quality elements according to the used quality measures for wall (hexahedral elements) and ILT (tetrahedral elements) models for the studied models.

| Part | Patient 1 Wall | ILT | Patient 2 Wall | ILT | Patient 3 Wall | ILT | Patient 4 Wall | ILT |
|---|---|---|---|---|---|---|---|---|
| No. of elements | 21,090 | 290,924 | 35,616 | 180,162 | 27,072 | 90,571 | 46,472 | 234,976 |
| No. of nodes | 116,883 | 251,506 | 197,160 | 309,160 | 149,648 | 168,802 | 256,780 | 391,362 |
| No. of elements failed to Jacobian | 0 | N/A | 0 | N/A | 5 | N/A | 0 | N/A |
| Min. Jacobian | 0.86 | N/A | 0.73 | N/A | 0.57 | N/A | 0.72 | N/A |
| No. of elements failed to volumetric skew | N/A | 3 | N/A | 0 | N/A | 84 | N/A | 147 |
| Max. vol. skew | N/A | 1 | N/A | 0.93 | N/A | 1 | N/A | 1 |
| No. of elements failed to min/max angle | 0 | 41 | 4 | 37 | 67 | 105 | 50 | 163 |
| Min angle | 50° | 0.5° | 45° | 9° | 26° | 6° | 37° | 5° |
| Max angle | 131° | 179° | 140° | 136° | 162° | 155° | 146° | 160° |

N/A: not applicable measure for the specific case.

### 3.3 Aneurysm wall stress

We extracted the maximum principal stress for the analysed case, as aneurysms will experience that highest stresses in the tangential direction which is consistent with the pressure vessel theory [39], Figure 10 shows the maximum principal stress contour plots. Table 3 compares the results of the peak and 99th percentile of maximum principal stress for the hexahedral meshes of aneurysm walls with results of tetrahedral meshes of aneurysm walls which include much larger number of nodes and elements (more than 1 million elements for an aneurysm [15]) for the same studied patients. Absolute differences showed that the results between the two computational grids are very close. The finite element models computation time for both grids is reported using Intel(R) Core(TM) i7-5930K CPU @ 3.50 GHz with 64.0 GB of RAM running Windows 8 OS, a significant time reduction is observed in the hexahedral meshes models.

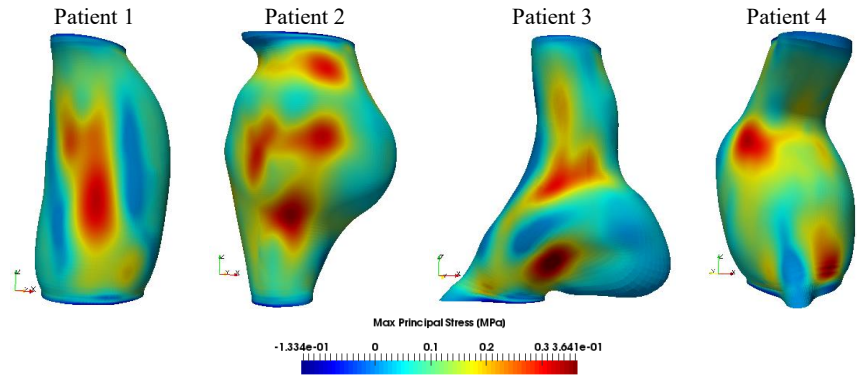

**Fig. 10** Maximum principal stress contour plots in the studied hexahedral meshes of aneurysm walls from the finite element solution.

**Table 3** Comparison of maximum principal results between hexahedral meshes and tetrahedral meshes of aneurysm walls, showing the peak and 99th percentile of the maximum principal stress for the four studied cases.

| | | Patient 1 | Patient 2 | Patient 3 | Patient 4 |
|---|---|---|---|---|---|
| Applied pressure/load (kPa) | | 12 | 13 | 13 | 14 |
| Peak of max. principal stress (MPa) | Hexahedral meshed wall | 0.3641 | 0.4543 | 0.4383 | 0.5048 |
| | Tetrahedral meshed wall | 0.3880 | 0.3953 | 0.4755 | 0.6074 |
| | Absolute difference | 0.0239 | 0.0590 | 0.0372 | 0.1026 |
| 99th percentile max. principal stress (MPa) | Hexahedral meshed wall | 0.2310 | 0.2859 | 0.3037 | 0.2437 |
| | Tetrahedral meshed wall | 0.2034 | 0.2554 | 0.2487 | 0.2176 |
| | Absolute difference | 0.0276 | 0.0305 | 0.0550 | 0.0261 |
| Computation time (sec) | Hexahedral meshed wall | 157 | 296 | 133 | 1355 |
| | Tetrahedral meshed wall | 880 | 1641 | 1196 | 1058 |

## 4    Discussion

Hexahedral meshing of aneurysm wall is not a trivial and straightforward problem. In this study, we present a technique to mesh the aneurysm wall using hexahedral elements using a commercial mesh generator, HyperMesh. The steps followed in this technique to get a high quality hexahedral meshes took $20 - 30$ minutes of manual work for each case. For future, these steps can be scripted using the language used by HyperMesh with minimum user-intervention. Low quality hexahedral elements could not be completely avoided (see Table 2, Patients 3 and 4) because of complex and irregular geometries of AAAs. Despite the fact that number of low quality elements did not exceed 0.2% of the total number of hexahedral elements in the patient-specific AAA model. From our experience mesh refinement would improve elements quality but will increase computational time with negligible change in results (2 hexahedral elements through wall thickness are sufficient for converged solution). As the key point in biomechanical analysis is to have an acceptable solution in reasonable time rather than optimal solution [32].

The created hexahedral meshes of aneurysm wall using ABAQUS/CAE (see 2.2.1) had high quality elements with minimum Jacobian ratio of 0.71. Although from Figure 11 the mesh density mismatch was obvious because of the inflexibility of defining partition planes in ABAQUS/CAE. Maximum principal stress contour plots showed very good match between the model created by ABAQUS/CAE meshing tool and the model created by HyperMesh, and the absolute difference of the peak value of maximum principal stress was 0.0479 MPa. HyperMesh was preferred on ABAQUS/CAE meshing tool as it requires less partitioning plans for mapped hexahedral meshes.

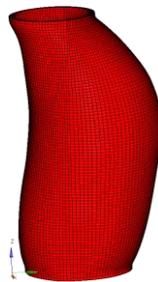

**Fig. 11** Hexahedral wall generated by ABAQUS/CAE meshing tool.

**Acknowledgments** The authors acknowledge funding of the Australian Government through the National Health and Medical Research Council NHMRC Ideas Grant, Ideas grant no. APP2001689. This research was carried out while the first author F. A. was in receipt of an "Australian Government Research Training Program Scholarship at The University of Western Australia". The first author acknowledges the support of Ms. Giuliana D'Aulerio of the University of Western Australia Medical School/ Division of Surgery for her contribution in obtaining the ethics approval. Contributions of Christopher Wood and Jane Polce, radiologists at Medical Imaging Department, Fiona Stanley Hospital, Murdoch, Western Australia in patient image acquisition are gratefully acknowledged.

# References

1.     Martufi, G. and T.C. Gasser, *the role of biomechanical modeling in the rupture risk assessment for abdominal aortic aneurysms.* Journal of biomechanical engineering, 2013. **135**(2): p. 021010.
2.     Johnston, K.W., et al., *Suggested standards for reporting on arterial aneurysms.* Journal of vascular surgery, 1991. **13**(3): p. 452-458.
3.     Wanhainen, A., et al., *Editor's choice–european society for vascular surgery (ESVS) 2019 clinical practice guidelines on the management of abdominal aorto-iliac Artery aneurysms.* European Journal of Vascular and Endovascular Surgery, 2019. **57**(1): p. 8-93.
4.     Upchurch, G.R. and T.A. Schaub, *Abdominal aortic aneurysm.* Am Fam Physician, 2006. **73**(7): p. 1198-204.
5.     Abubakar, I., T. Tillmann, and A. Banerjee, *Global, regional, and national age-sex specific all-cause and cause-specific mortality for 240 causes of death, 1990-2013: a systematic analysis for the Global Burden of Disease Study 2013.* Lancet, 2015. **385**(9963): p. 117-171.

6.      Lancet-2017. *The lancet, Global Burden of Disease*. April 2022]; Available from: https://www.thelancet.com/gbd/gbd-compare-visualisation.

7.      Scott, R. and M.A.S.S. Group, *The Multicentre Aneurysm Screening Study (MASS) into the effect of abdominal aortic aneurysm screening on mortality in men: a randomised controlled trial.* The Lancet, 2002. **360**(9345): p. 1531-1539.

8.      Powell, J.T. and A.R. Brady, *Detection, management, and prospects for the medical treatment of small abdominal aortic aneurysms.* Arteriosclerosis, thrombosis, and vascular biology, 2004. **24**(2): p. 241-245.

9.      Geest, J.V., et al., *A biomechanics-based rupture potential index for abdominal aortic aneurysm risk assessment: demonstrative application.* Annals of the New York Academy of Sciences, 2006. **1085**: p. 11-21.

10.     Gasser, T.C., et al., *Biomechanical rupture risk assessment of abdominal aortic aneurysms: model complexity versus predictability of finite element simulations.* European Journal of Vascular and Endovascular Surgery, 2010. **40**(2): p. 176-185.

11.     Gasser, T.C., et al., *A novel strategy to translate the biomechanical rupture risk of abdominal aortic aneurysms to their equivalent diameter risk: method and retrospective validation.* European Journal of Vascular and Endovascular Surgery, 2014. **47**(3): p. 288-295.

12.     Polzer, S. and T.C. Gasser, *Biomechanical rupture risk assessment of abdominal aortic aneurysms based on a novel probabilistic rupture risk index.* Journal of The Royal Society Interface, 2015. **12**(113): p. 20150852.

13.     Joldes, G.R., et al., *BioPARR: A software system for estimating the rupture potential index for abdominal aortic aneurysms.* Scientific reports, 2017. **7**(1): p. 4641.

14.     Raut, S.S., P. Liu, and E.A. Finol, *An approach for patient-specific multi-domain vascular mesh generation featuring spatially varying wall thickness modeling.* Journal of biomechanics, 2015. **48**(10): p. 1972-1981.

15.     Miller, K., et al., *Is There a Relationship Between Stress in Walls of Abdominal Aortic Aneurysm and Symptoms?* Journal of Surgical Research, 2020. **252**: p. 37-46.

16.     Jessica, Z.Y., *Image-Based Quadrilateral and Hexahedral Meshing*, in *Geometric Modeling and Mesh Generation from Scanned Images*. 2018. p. 193-228.

17.     Wittek, A., et al., *On stress in abdominal aortic aneurysm: Linear versus non-linear analysis and aneurysm rupture risk.* International Journal for Numerical Methods in Biomedical Engineering, 2022. **38**(2): p. e3554.

18.     De Santis, G., et al., *Patient-specific computational fluid dynamics: structured mesh generation from coronary angiography.* Medical & biological engineering & computing, 2010. **48**(4): p. 371-380.

19.     Trachet, B., et al., *An integrated framework to quantitatively link mouse-specific hemodynamics to aneurysm formation in angiotensin II-infused ApoE−/− mice.* Annals of biomedical engineering, 2011. **39**(9): p. 2430-2444.

20.     Marchandise, E., C. Geuzaine, and J.-F. Remacle, *Cardiovascular and lung mesh generation based on centerlines.* International journal for numerical methods in biomedical engineering, 2013. **29**(6): p. 665-682.

21.     Tarjuelo-Gutierrez, J., et al., *High-quality conforming hexahedral meshes of patient-specific abdominal aortic aneurysms including their intraluminal thrombi.* Medical & biological engineering & computing, 2014. **52**(2): p. 159-168.

22.     Joldes, G.R., et al., *A simple method of incorporating the effect of the Uniform Stress Hypothesis in arterial wall stress computations.* Acta of Bioengineering and Biomechanics,, 2018. **20**(3): p. 59-67.

23.     Mayr, M., W.A. Wall, and M.W. Gee, *Adaptive time stepping for fluid-structure interaction solvers.* Finite Elements in Analysis and Design, 2018. **141**: p. 55-69.

24.     Auer, M. and T.C. Gasser, *Reconstruction and finite element mesh generation of abdominal aortic aneurysms from computerized tomography angiography data with minimal user interactions.* IEEE transactions on medical imaging, 2010. **29**(4): p. 1022-1028.

25.     Polzer, S., et al., *A numerical implementation to predict residual strains from the homogeneous stress hypothesis with application to abdominal aortic aneurysms.* Annals of biomedical engineering, 2013. **41**(7): p. 1516-1527.

26. Zhang, Y. and C. Bajaj, *Adaptive and quality quadrilateral/hexahedral meshing from volumetric data.* Computer methods in applied mechanics and engineering, 2006. **195**(9-12): p. 942-960.

27. Maier, A., et al., *A comparison of diameter, wall stress, and rupture potential index for abdominal aortic aneurysm rupture risk prediction.* Annals of biomedical engineering, 2010. **38**(10): p. 3124-3134.

28. Fedorov, A., et al., *3D Slicer as an image computing platform for the Quantitative Imaging Network.* Magnetic resonance imaging, 2012. **30**(9): p. 1323-1341.

29. Andy T. Huynh, K.M., *Towards accurate measurement of abdominal aortic aneurysm wall thickness from CT and MRI*, in *Computational Biomechanics for Medicine - Towards translation and better patient outcomes*. 2022, Springer International Publishing.

30. Bern, M.W. and P.E. Plassmann, *Mesh Generation.* Handbook of computational geometry, 2000. **38**.

31. Ito, Y., A.M. Shih, and B.K. Soni, *Octree-based reasonable-quality hexahedral mesh generation using a new set of refinement templates.* International Journal for Numerical Methods in Engineering, 2009. **77**(13): p. 1809-1833.

32. Yang, K.-H., *Basic finite element method as applied to injury biomechanics*. 2017: Academic Press.

33. Smith, M., *ABAQUS/Standard User's Manual, Version 6.9.* Providence, RI: Simulia, 2009.

34. Joldes, G.R., et al., *A simple, effective and clinically applicable method to compute abdominal aortic aneurysm wall stress.* Journal of the Mechanical Behavior of Biomedical Materials,, 2016. **58**: p. 139-148.

35. Raghavan, M., B. Ma, and M.F. Fillinger, *Non-invasive determination of zero-pressure geometry of arterial aneurysms.* Annals of biomedical engineering, 2006. **34**(9): p. 1414-1419.

36. Riveros, F., et al., *A pull-back algorithm to determine the unloaded vascular geometry in anisotropic hyperelastic AAA passive mechanics.* Annals of biomedical engineering, 2013. **41**(4): p. 694-708.

37. Shigley, J.E., *Shigley's mechanical engineering design*. 2011: Tata McGraw-Hill Education.

38. Speelman, L., et al., *Patient-specific AAA wall stress analysis: 99-percentile versus peak stress.* European Journal of Vascular and Endovascular Surgery, 2008. **36**(6): p. 668-676.

39. Budynas, R.G., *Advanced strength and applied stress analysis*. 1999: McGraw-Hill.